# Computational Essays in the Physics Classroom

Tor Ole B. Odden and John Burk

**Introduction: Argumentation and programming in physics**

Writing and argumentation are critical to both professional physics and physics education. They are both necessary to the practices of professional science like paper writing and peer review, and are increasingly emphasized in science education standards and best practices [1,2]. Despite this, the skill of making an extended argument in writing is often overlooked in physics classrooms, apart from certain isolated practices like lab notebooks or mathematical proofs.

Computation is also critical to both professional physics and, increasingly, physics education [3]. Over the past 30 years computation has grown into a common, and often indispensable, part of the physics classroom. Visual, interactive tools like PhET simulations, Easy Java Simulations, and VPython [4–6], have steadily lowered the costs and barriers to entry for computational modelling and recent pushes to expand students' opportunities to learn programming [7] have only added momentum to this trend. Now, with the advent of new computational technology we can use computation to facilitate writing and argumentation in physics, through the use of what are known as computational essays.

**What is a computational essay?**

The term *computational essay* was introduced by Andrea diSessa in his book *Changing Minds* (diSessa, 2000) as a type of essay that uses text, along with small programs, interactive diagrams, and computational tools to express an idea. Stephen Wolfram, the chief designer of Mathematica, has also discussed computational essays as documents that use text, computer input, and computer output to explore and communicate ideas [9]. Computational essays are, in short, pieces of writing that explicitly incorporate code to support their theses. They include all of the elements one would expect in an ordinary essay (introduction, thesis statement, body paragraphs, conclusion) and have a similar set of goals (to present a step-by-step argument or explanation). However, the argument in a computational essay is driven by the output of various blocks of code, with the text serving to make clear the function of different code blocks and to describe the output (see Figures 1 and 2).

Because they involve both text and computer code, computational essay must be written in environments that allow you to mix the two. One example of this is embedding small interactive programs, like Glowscript or Python Trinkets, into blogs or web pages (for an example of this see Allain, 2017). Another set of commonly-used environments are so-called *notebooks*. Notebooks are programming environments that consist of "inputs" into which users can enter code and text, as well as typeset equations, images, and videos. Each input can hold as much code or text as the user desires, from single lines to whole programs or paragraphs.

Code inputs can be run in any order, or even multiple times if the programmer prefers, allowing them to test small-scale changes in a program without re-running the entire script. There are several types of notebook software currently available. Many readers will likely be familiar with *Mathematica*, a notebook-based mathematical tool that has been available, under license, for many years [11]. However, in recent years a free type of notebook software, known as the *Jupyter notebook*, has been developed which allows users this same type of flexibility with Python and other commonly-used programming languages [12]. Jupyter notebooks are free to download and use, and are increasingly used by professional data scientists, engineers, and physicists to present their work [13].

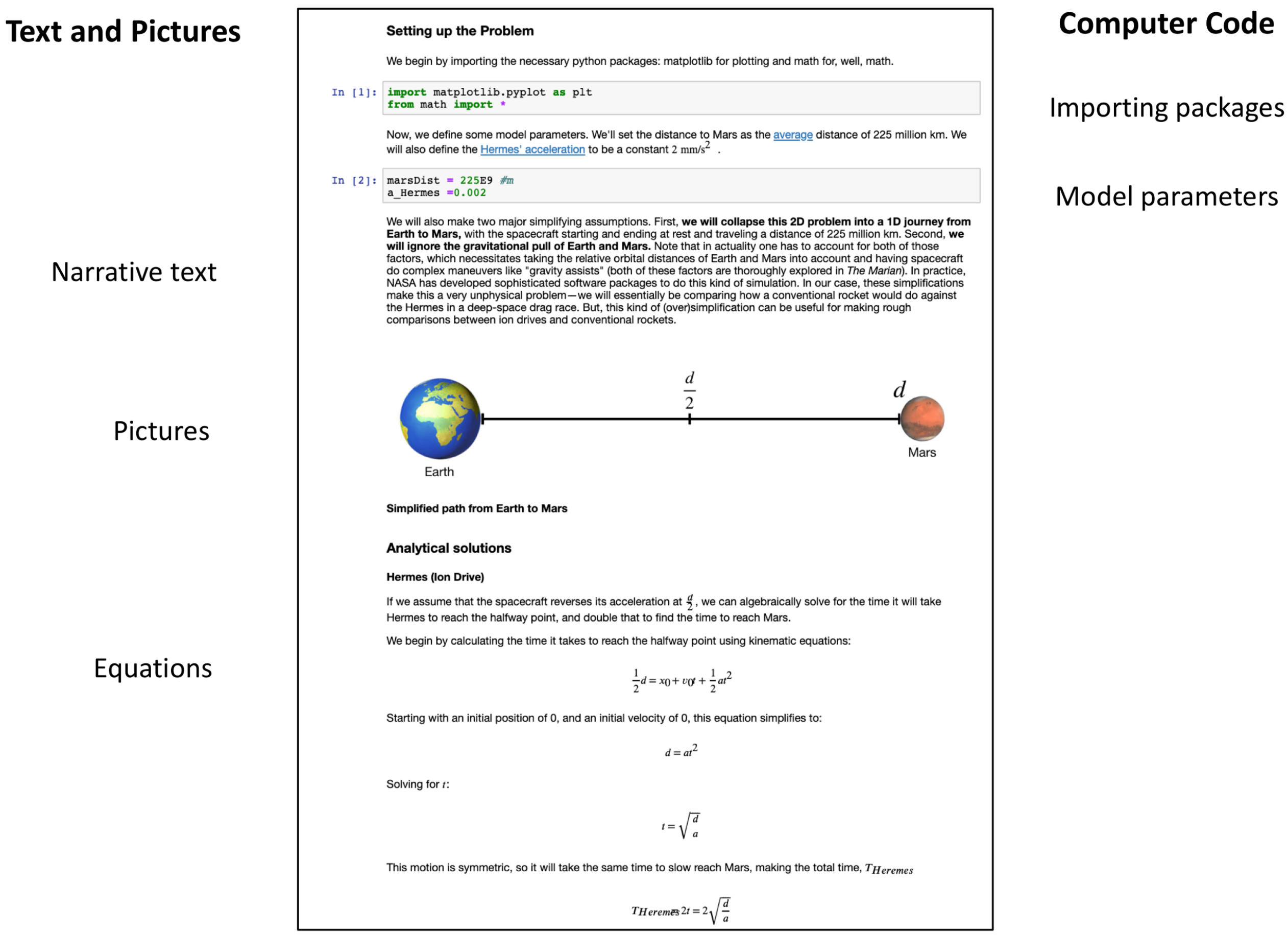

**Figure 1:** An example computational essay in a Jupyter notebook on the topic of ion drives vs. conventional rockets using the rocket equation at constant thrust. The essay is available at https://uio-ccse.github.io/computational-essay-showroom/

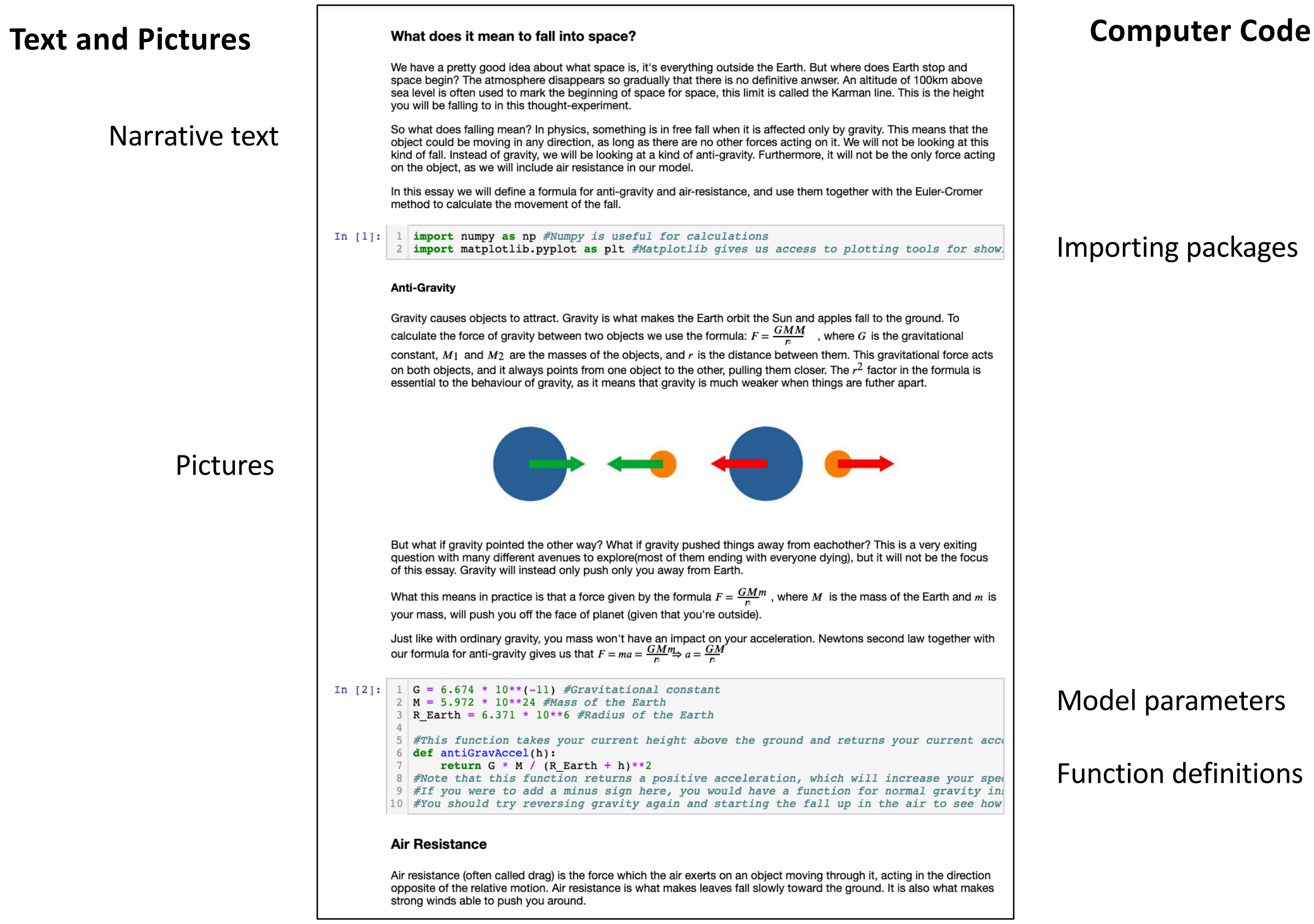

**Figure 2:** An example computational essay in a Jupyter notebook on the topic of how long it would take to fall into space (assuming gravity reversed) using atmospheric air resistance. The essay is available at https://uio-ccse.github.io/computational-essay-showroom/

**Computational essays in physics education**

Computational essays hold great potential for physics educators. At a basic level, they can simplify (or eliminate!) many aspects of programming in education with which we often struggle. For example, in cases where a student is asked to complete a programming project and accompanying report, both the code and report can be combined into a single document, avoiding the hassle of comparing two different writeups. Teachers can also use computational essays to scaffold longer computational assignments or investigations, for example guiding students through the development and elaboration of a computational model. Additionally, teachers can use them to introduce students to new computational or modeling techniques by creating multi-step, scaffolded, interactive tutorials.

However, beyond these basic applications, we see computational essays as being potentially powerful structures for scientific communication, argumentation, and programming. Because the code in a computational essay is visible and has to be incorporated into the argument, students build the skills of communicating about their code and explaining their investigations. This type of communication is a key scientific practice [1] and computational essays give students the opportunity to practice it in contexts beyond standard lab write-ups or reports. Additionally, because the code can be executed and modified by readers, computational essay investigations are inherently replicable. In other words, the code in a computational essay serves as both the method and results of an investigation, and because it is directly integrated into the document readers (both teachers and other students) can easily verify that a student's program does indeed work as advertised.

Computational essays are also powerful exploratory tools. Because the code in a computational essay is "live" and interactive, students can execute and re-execute their code as they compose their investigation. This makes it easy to "play around" with a program and explore the effects of minor changes on a simulation, potentially uncovering new factors that might require a student to adjust, complexify, or rework their argument. This experience of reworking an argument in light of new discoveries is common for professional scientists but not often experienced by students. A computational essay "reader" can also make these kinds of changes, allowing students to critique each other's work while also testing the theories and questions they may come up with when reading another student's essay.

**What does this look like in practice?**

At the University of Oslo we are starting build computational essays into the work requirements for our introductory physics courses. Since the early 2000s programming has been a core part of the University of Oslo physics curriculum, with all physics majors taking both a programming course and a numerical methods course during their first semester. However, until recently programming has mostly been used for small-scale assignments in subsequent courses (such as simulating a runner with air resistance), with little focus on writing or communication. During the fall semester of 2018, we had a small pilot group of students write computational essays in their introductory electricity and magnetism course, as an alternative to a mandatory presentation-based project that all students had to complete. 17 students participated in this pilot, working either individually or in pairs.

Participating students received the following challenge: *use a simulation to investigate a problem that you find interesting, then write a computational essay about what you've learned.* This deliberately left the task open in the hopes that the students' coding and investigations would be driven by their own interests rather than prescribed by us. We also provided the students with several forms of scaffolding and support for their projects. First and foremost, this included a set of basic pre-built simulations, written in Jupyter notebooks, for the students to build off of. These simulations were based on the electricity and magnetism topics covered in the course and featured phenomena like cyclotrons, storm clouds and lightning, railguns, and magnetic traps [14]. However, the simulations were deliberately simplified, neglecting many important physical factors, to give students room for exploration. We also provided the students with an example computational essay to set the level of expectation for the project, and held twice-weekly drop-in sessions where they could go for help with any physics, coding, or logistical questions. Students were given roughly 4-6 weeks to work on their essays. Once the students finished, we organized presentations where participants presented their work to a small group of their peers by putting their essays up on a projector and walking the group through their investigation. This practice was modeled after the way that actual computational physics research groups present results to one another.

In the end, all students in the pilot group successfully completed computational essays. Students reported spending around 6-14 hours on the project, and chose a roughly even spread of topics, with no two groups choosing exactly the same investigation question. For example, one pair of students investigated whether a railgun could be a viable mechanism to power the metro system in the city of Oslo. Others looked at different ways magnetic traps could be used to confine charged particles, the conditions under which one would remain safe near a lightning strike, and the effects of relativity on particles in a particle accelerator. Students made varying levels of modification to the given code, ranging from taking the existing code and running it multiple times to re-writing it entirely. Students also spent significant time reading external websites and papers for background information on their phenomena, simulation parameters, and published results for comparison.

In follow-up interviews, students expressed appreciation for the creative freedom and non-linear nature of the project. For example, one student favorably compared computational essays to standard computational assignments, saying

> *I think we pushed ourselves harder here than we would with those assignments. Because then you have an endpoint like, okay, I've done what the program or what the assignment asked me to do and here's the program. But now when we finished something it was like 'this is really cool to actually see. What else can we do?'*

However, students also reported that this openness was challenging in that it could be difficult to choose a topic and investigation question. Some students also had difficulties with the non-linear nature of notebooks, where stored variables could be carried over between different blocks of code, and many wished for more time to build out their projects. To address these challenges, in future iterations of the project we plan to provide the students with additional time to work on their projects, training in notebook use, and suggested topics for investigation.

**Getting started with computational essays**

Based on these preliminary experiences, we see great potential for computational essays in physics education, both in high school and college settings. In courses that already use programming, computational essays could add another class of assignment, which could be used for both teaching and assessment. For example, computational essays could act as large-scale projects that students would work on over the course of a semester, with repeated feedback and rounds of revision. Multiple computational essays could then form the basis of a portfolio-based assessment. For courses that do not yet use or emphasize programming these essays could be a stepping stone to programming integration, allowing students to see and play around with small programs framed in a larger narrative. A good starting point for interested readers and educators is the work of Rhett Alain at the Dot-Physics section of Wired [10] and the computational essay repository at the University of Oslo [15].

We also see a great need for the development of educational materials using these tools. For example, to successfully write computational essays most students will need explicit guidance on explicit guidance on effective practices. However, although there are commonly-taught guidelines for ordinary essays such guidelines do not yet exist for computational essays. We also see a need for additional examples of well-written computational essays at a variety of levels. At the University of Oslo we have begun to put together such examples, based on the work of both researchers and students, but we look forward to seeing many more as computational essays make their way into the physics education community.

**Acknowledgments**
This project was funded by NOKUT, the Norwegian Agency for Quality Assurance in Education, and the Research Council of Norway. The authors would like to thank Danny Caballero, Anders Malthe-Sørenssen, John Mark Aiken, Henrik Sveinsson, and Karl Henrik Fredly for their help and support in this project.